# $Ca_{25}Co_{22}O_{56}(OH)_{28}$: a layered misfit compound

*T. Klimczuk[1,2], H.W. Zandbergen[1,3], N. M. van der Pers[4], L. Viciu[1], V.L. Miller[1], M.-H. Lee[5], and R.J. Cava[1]*

[1]Department of Chemistry, Princeton University, Princeton NJ 08544,

[2]Faculty of Applied Physics and Mathematics, Gdansk University of Technology, Narutowicza 11/12, 80-952 Gdansk, Poland,

[3]National Center for HREM, Department of Nanoscience, Delft University of Technology, Rotterdamseweg 137, 2682 AL Delft, The Netherlands

[4]Department of Materials Science, Delft University of Technology, Rotterdamseweg 137, 2682 AL Delft, The Netherlands

[5]Department of Physics, Princeton University, Princeton NJ 08544,

**Abstract**

The high pressure synthesis, structure and magnetic properties of $Ca_{25}Co_{22}O_{56}(OH)_{28}$ are reported. The compound has a misfit structure, consisting of double, square calcium oxide hydroxide rock-salt-like layers between hexagonal $CoO_2$ layers. The misfit compound crystallizes in the monoclinic space group C2/m, and can be characterized by the coexistence of two subsystems with common a=4.893(5) Å, c=8.825(9) Å and β=95.745(8) parameters, and different *b* parameters: $b_{RS}$=4.894(5) Å, and $b_{HEX}$=2.809(3) Å, for the rock-salt and hexagonal type planes respectively. The compound shows Curie-Weiss paramagnetism with an antiferromagnetic Weiss temperature of -43K and a reduced Co moment. Substantial deviations from Curie-Weiss behavior are seen below 50K with no indication of magnetic ordering. No superconductivity was observed down to a temperature of 2K.

**Introduction**

The observations of an anomalously high thermoelectric effect in $Na_{0.7}CoO_2$ and superconductivity in $Na_{0.3}CoO_2 \cdot 1.3H_2O$ have resulted in substantial interest in the properties of layered materials based on layers of edge shared $CoO_6$ octahedra [1, 2]. These triangular $CoO_2$ layers are the central structural building block of a whole family of recently elucidated compounds. In these compounds, the $CoO_2$ layers are stacked with square AO layers of the NaCl-type [3-10]. The triangular and square layers do not fit perfectly, making the compounds of the misfit type, similar to that observed previously for chalcogenides [11-13]. The equal lengths of the [100] direction in the square layers and the [120] direction in the triangular layers allow them to fit in one in-plane direction while misfitting along another. The relative periodicity of the layers determines the overall composition, as $(AO)_n(CoO_2)_m$.

There are three previously known compounds in the Ca-Co-O system: $Ca_3Co_2O_6$ with one dimensional chains built by face-sharing $CoO_6$ polyhedra [14], $Ca_xCoO_2$, with alternating hexagonal $CoO_2$ and Ca layers [15], and $Ca_3Co_4O_9$, a misfit compound consisting of $CoO_2$ layers stacked with CaO–CoO–CaO triple rock salt layers [4]. In $Ca_3Co_4O_9$, Co atoms are present not only in the $CoO_2$ layers but also in a Co-O rock salt layer. We postulated that the presence of Co in the rock salt layer results in magnetic coupling that likely disrupts the layered nature of the magnetic system. The goal of the current study was to synthesize a Ca-Co-O misfit structure with no Co-O layer in the rocksalt portion of the structure. High pressure synthesis was employed to achieve this goal. The compound reported here is essentially isostructural with one recently reported [16], but differs significantly in composition and properties, as described below.

**Experiment**

Polycrystalline Ca-Co-O-OH samples were synthesized under high pressure in the following fashion. The precursors $Ca(OH)_2$ (made by room temperature hydration of CaO, Puratronic 99.998%) and $Co_3O_4$ (Alfa Aesar 99.7%) were weighed in the appropriate proportions, and 25 wt. % of an oxidizer ($NaClO_3$) was added. For some syntheses, $Y_2O_3$ was employed as well. The thoroughly mixed powder was sealed in a gold capsule (5mm diameter) and pressed in a cubic multi-anvil module in a commercial high-pressure furnace (Rockland Research). The samples were heated under a pressure of 4.5 GPa, with a heating rate of 50°C/min, reacted at 800°C for 3 hours, and then quenched under pressure to room temperature. The final product was ground and washed in water to remove the NaCl present due to the decomposition of the oxidizer. The new phase did not appear in any detectable quantity when the gold capsule used for the synthesis was sealed in a dry box with dry $Co_3O_4$ and dry CaO used as starting materials. In repeated syntheses, only with the use of calcium hydroxide could the new compound be obtained.

X-ray powder diffraction (XRD) was performed with $CuK_\alpha$ radiation on a diffractometer equipped with a diffracted beam monochromater. Electron microscopy was performed on electron transparent areas that were obtained by crushing and dropping a suspension of the powder in ethanol on a carbon coated holey film. High resolution electron microscopy (HREM) and electron diffraction were performed with a Philips CM200ST electron microscope with a field emission gun operated at 200 kV and equipped with an EDX (Energy dispersive X-ray) element analysis system. Thermogravimetric analysis (TGA) was carried out in a commercial apparatus (Perkin Elmer). The magnetic susceptibility was measured on powder sample with a SQUID magnetometer (MPMS, Quantum Design) in the temperature range 2-350K.

**Results**

A series of samples was made with a variety of whole number ratios near 1:1, expected to be close to the composition of a "misfit" phase with two CaO layers inserted between single $CoO_2$ layers. Analysis of the powder X-ray diffraction patterns showed the presence of a new phase that appeared only in the high pressure syntheses. Electron diffraction analysis indicated that the new phase is of the misfit type. A characteristic electron diffraction pattern is shown in figure 1. Using the relative periodicities of the rock salt and hexagonal structures that are clearly revealed in the diffraction pattern, the ratio of the rocksalt part (CaO-OH) to the hexagonal part ($CoO_2$) of the misfit structure can be determined. Figure 2a shows two planes in the misfit structure of the new phase that best illustrate the character of the misfit: one rocksalt plane and one oxygen plane in the $CoO_2$ layer are shown. The square rocksalt unit cell with a and $b_{RS}$ is shown in the upper panel, and the hexagonal unit cell of the $CoO_2$ layer with $b_{HEX}$ is shown in the lower panel. This figure is drawn with the periodicities of the two layers to match what is seen in the electron diffraction patterns: a good match in the "a" direction, but, in the b direction, only after 25 units of rocksalt and 22 of hexagonal $CoO_2$ are both lattices in register. Simple geometrical calculations from the electron diffraction pattern parameters $b_{RS}$=4.90 Å (having 4 Ca atoms per unit cell) and $b_{HEX}$=2.81 (having 2 Co atoms per unit cell) gives a Ca:Co ratio of 1.146. This yields a more exact formula, if both layers were only oxide, of $(CaO)_{25.4}(CoO_2)_{22}$, which we approximate as $(CaO)_{25}(CoO_2)_{22}$. Material synthesized by the method described, with the formula determined by analysis of the electron diffraction data, 25Ca:22Co, resulted in a pure phase.

The oxygen and hydroxide content of the pure phase material was determined by heating in flowing $H_2$ in a TGA at a rate of 0.25 degree per minute to a temperature of 700 °C. The data are shown in figure 3. The final products of the reduction were found by X-ray diffraction to be Co metal and CaO. The figure shows that there are two distinct temperatures where weight loss occurs on heating. The first weight loss on heating occurs near 300°C - the same temperature where the OH is lost from the related compounds $Sr_3Co_2O_5(OH)_{1.2} \cdot 1.2H_2O$, $Sr_3NdFe_3O_{7.5}(OH)_2 \cdot xH_2O$, and $Sr_4Co_{1.6}Ti_{1.4}O_8(OH)_2 \cdot xH_2O$ [17,18,19]. This first weight loss is much higher in temperature than where water is lost from the related Sr-Co(Fe)-O-OH-$H_2O$ compounds or other hydrates (usually near 100 °C). The second weight loss is from the loss of oxygen. Analysis of the two weight loss amounts indicates that the formula of the present compound is $Ca_{25}Co_{22}O_x(OH)_y$ where x is between 56 and 57 and y is between 28 and 27, with the uncertainly arising from experimental accuracy. We conclude that the formula of this complex material is best described as $Ca_{25}Co_{22}O_{56}(OH)_{28}$; the alternative, $Ca_{25}Co_{22}O_{57}(OH)_{27}$, would yield a formal oxidation for Co that is too high. The formal oxidation state for Co in $Ca_{25}Co_{22}O_{56}(OH)_{28}$ is $Co^{+4}$. The present material has the same basic structure but is significantly different from recently reported $(CaOH)_{1.14}CoO_2$, for which the Co oxidation state is 2.9+ [16]. The reduced oxidation state for Co in $(CaOH)_{1.14}CoO_2$ is consistent with the higher temperature (1100-1200°C), more reducing atmosphere (no oxidizer was included), and lower OH content employed in the synthesis of that compound.

Through a combination of the XRD and ED information, a more quantitative model for the structure of $Ca_{25}Co_{22}O_{56}(OH)_{28}$ can be proposed. The powder X-ray diffraction pattern, figure 4, shows the coexistence of two monoclinic subsystems with common a, c, and $\beta$ parameters and different b parameters: $b_{RS}$ and $b_{HEX}$ for rock-salt and hexagonal type planes respectively [4]. $Ca_{25}Co_{22}O_{56}(OH)_{28}$ can be described in the monoclinic system (space group C2/m): and can be characterized by the coexistence of two subsystems with common a=4.893(5) Å, c=8.825(9) Å and β=95.745(8) parameters, and different b parameters: $b_{RS}$=4.894(5) Å, and $b_{HEX}$=2.809(3) Å, for the rock-salt and triangular planes respectively. The analysis of the X-ray powder diffraction data from figure 4 is summarized in table 1, with Miller indices assigned by this misfit system. The cell parameters for the new phase are consistent with what has been previously reported for $Ca_3Co_4O_9$, which contains triple rock salt layers (CaO–CoO–CaO ) [4, 9, 10], with the intermediate CoO layer absent. The smaller $b_{hex}$ and larger c are consistent with an increased Co oxidation state and increased O and OH content of the current material when compared to $(CaOH)_{1.14}CoO_2$. The model for the structure is shown in figure 2b.

Detailed structural refinement of misfit compound structures is beyond the current study. However, the *00l* reflections of the diffraction pattern can easily be employed to refine the positions of the CaO-CaO-CoO$_2$ layers in the structure (The OH was not included). The intensities of such reflections are sensitive only to the heights of the layers along the stacking axis, and the scattering power in each layer. Our initial model for the structure was derived by considering the distance between the Co and the O layers in the triangular CoO$_2$ layer in Na$_{0.3}$CoO$_2$, and the distance between 100 type CaO layers in rock salt structure CaO. Five *00l* reflections were clearly observable in the diffraction pattern. Two structural parameters were refined: the "z"s (fractional heights) of the O layers around the Co layer, and the CaO layers (the O layers were assumed to be spaced equally above and below the Co layer; as were the CaO layers.) The intensities of the reflections were fit very well (The agreement for the 00*l* peak integrated intensities ($R_{Bragg}$) is 2.6%) through refining the two free z parameters by least squares, as shown in table 2. The refinements are extremely sensitive to the number of metal layers in the structure, and to mixing of Co into the CaO layers, and thus the model for the crystal structure of $Ca_{25}Co_{22}O_{56}(OH)_{28}$ is confirmed by quantitative analysis of the diffraction data. The distances between the planes as determined in the refinements are shown in table 2. Though no information can be obtained about the distribution of OH in the compound from our data, we speculate that the OH molecules are most likely bonded to Ca, primarily accommodated in the misfit region between the hexagonal and square layers.

For superconductivity in layered hexagonal cobaltates, only one example, Na$_{0.3}$CoO$_2 \cdot$1.3H$_2$O ($T_C$=4.5K) is presently known [2]. Assuming that only the Na content determines the Co valency, a formal oxidation state of $Co^{+3.7}$ appears to be preferred for superconductivity. The oxidation state for Co in the present compound is 4 and therefore superconductivity might not occur due to the presence of an inappropriate charge state. We therefore tried to lower the formal oxidation state to 3.7 by attempting to synthesize $Y_xCa_{25-x}Co_{22}O_{56}(OH)_{28}$ with x=6.6. However, the X-ray diffraction study performed on that sample showed that both YCoO$_3$ and Y$_2$O$_3$ phases were present in substantial quantities. This indicated that x=6.6 substantially exceeds the Y solubility limit in $Y_xCa_{25-x}Co_{22}O_{56}(OH)_{28}$. Although some Y was accommodated in the new phase, the resulting material was not superconducting above 2K.

The zero field cooled (ZFC) DC magnetization data for $Ca_{25}Co_{22}O_{56}(OH)_{28}$, measured on heating under an applied magnetic field of 10 kOe, are presented in figure 6, which shows the inverse of magnetic susceptibility between 5 and 350K. Fitting the high-temperature data (50-350K) to the Curie-Weiss law, including the temperature independent part of the susceptibility $\chi_0$ (Pauli paramagnetism), yielded a Weiss temperature ($\theta$) of -43K, an effective magnetic moment $\mu_{eff.}$=1.1$\mu_B$/mol-Co, and $\chi_0$=1.2 x $10^{-4}$ emu/Oe/mol-Co. The effective moment per Co corresponds to less than a full spin 1/2 $Co^{+4}$ per formula unit, suggesting that the spin is not fully localized. The significant Pauli paramagnetism suggests that the material may be a metallic conductor, but the presence of NaCl between all grains in the synthesized samples does not allow for the measurement of intrinsic resistivities. Figure 5 shows that the material displays a substantial deviation from the Curie-Weiss law at temperatures below 50K. The susceptibility varies smoothly and is below the extrapolation of the high temperature Curie Weiss fit, suggesting that ferromagnetic fluctuations are becoming more significant in the magnetic system at low temperature. This behavior is often observed in frustrated spin systems, as expected for a triangle-based Co-O lattice. The magnetic susceptibility for the current material is substantially higher, with a larger local moment and larger Weiss theta than has been reported [16] for $(CaOH)_{1.14}CoO_2$, consistent with the presence of a significant fraction of localized, spin ½, $Co^{4+}$. No indication of magnetic ordering or tendency toward magnetic ordering is seen in the susceptibility down to temperatures of 2K. The behavior observed in this material is very similar to what is seen for $Na_{0.7}CoO_2$ [21].

**Conclusions**

Cobalt oxides based on hexagonal $CoO_2$ layers are beginning to reveal an interesting new class of structure-property relations not generally observed in materials based on square lattices. Though this family of materials is not as extensive as the perovskites, the compounds recently discovered indicate that there may be substantial opportunity to develop them farther in the future. The synthetic experiments described here suggest that high pressure may be important in the development of that chemistry.

**Acknowledgements**

This research was supported by the US Department of Energy, Basic Energy Sciences, grant DE FG02-98-ER45706, and the NSF MRSEC and Solid State Chemistry programs, grant numbers DMR-0213706 and DMR-0244254. T.K. acknowledges support from the Foundation for Polish Science.

**Table 1.** Powder X-ray diffraction pattern of $Ca_{25}Co_{22}O_{56}(OH)_{28}$ (Cu $K_\alpha$ radiation). With refined cell parameters for the two subsystems:

$a = 4.893(5)$, $c_1 = 8.825(9)$, $\beta = 95.745(8)$, $b_{RS} = 4.894(5)$, $b_{HEX} = 2.809(3)$ Å

| $2\Theta_{exp}$ (deg) | Int. | $h_1$ | $k_1$ | $l_1$ | $2\Theta_{cal}$ (deg) | $h_2$ | $k_2$ | $l_2$ | $2\Theta_{cal}$ (deg) |
|---|---|---|---|---|---|---|---|---|---|
| 10.12 | 1000 | 0 | 0 | 1 | 10.06 | 0 | 0 | 1 | 10.06 |
| 20.28 | 459 | 0 | 0 | 2 | 20.21 | 0 | 0 | 2 | 20.21 |
| 30.59 | 631 | 0 | 0 | 3 | 30.51 | 0 | 0 | 3 | 30.51 |
| 36.97 | 9 | 2 | 0 | 0 | 36.90 | 1 | 1 | 0 | 36.91 |
| 38.92 | 6 | 1 | 1 | -3 | 38.85 | 1 | 1 | 1 | 38.86 |
| 40.57 | 6 | 2 | 0 | -2 | 40.50 | 2 | 0 | -2 | 40.49 |
| 41.16 | 21 | 0 | 0 | 4 | 41.08 | 0 | 0 | 4 | 41.08 |
| 43.47 | 8 | | | | | 1 | 1 | 2 | 43.37 |
| 46.08 | 10 | 2 | 0 | -3 | 46.00 | 2 | 0 | -3 | 46.00 |
| 47.44 | 4 | 1 | 1 | -4 | 47.42 | 1 | 1 | -3 | 47.33 |
| 49.93 | 6 | | | | | 1 | 1 | 3 | 49.87 |
| 51.16 | 5 | 2 | 0 | 3 | 51.11 | 2 | 0 | 3 | 51.09 |
| 52.10 | 22 | 0 | 0 | 5 | 52.03 | 0 | 0 | 5 | 52.02 |
| 54.92 | 5 | 2 | 2 | 1 | 54.85 | 1 | 1 | -4 | 54.86 |
| 57.13 | 6 | 1 | 1 | -5 | 57.06 | | | | |
| 57.99 | 6 | | | | | 1 | 1 | 4 | 57.91 |

**Table 2:** *00l* Reflection Refinement[1] of CaO and CaO$_2$ Layer Heights in Ca$_{25}$Co$_{22}$O$_{56}$(OH)$_{28}$

Structural Parameters:

*c* axis length: 8.7773(3) Å

| Atoms[2] | Planes/Cell | Occupancy | z |
|---|---|---|---|
| CaO plane | 2 | 0.573 | 0.6393(3) |
| O plane | 2 | 1.000 | 0.913(1) |
| Co | 1 | 1.000 | 0.0000 |

Distances Å

| | |
|---|---|
| CoO$_2$ layer thickness: | 1.52 |
| CaO-CaO layer separation | 2.44 |
| CaO plane to O plane in CoO$_2$ layer | 2.41 |

Calculated and observed relative intensities

| h | k | l | 2θ° | d-spacing Å | Icalc. | Iobs. |
|---|---|---|---|---|---|---|
| 0 | 0 | 1 | 10.070 | 8.77727 | 972 | 1000 |
| 0 | 0 | 2 | 20.218 | 4.38863 | 446 | 455 |
| 0 | 0 | 3 | 30.530 | 2.92576 | 593 | 610 |
| 0 | 0 | 4 | 41.102 | 2.19432 | 17 | 19 |
| 0 | 0 | 5 | 52.055 | 1.75545 | 28 | 27 |

Agreement (R Bragg)     2.61%

[1]Cu Kα radiation, diffracted beam monochromater. Included regions (degrees 2θ): 9-11.5 (001), 19-21.5 (002), 29.5- 31.5 (003) 40-43 (004), 51-53(005). Background refined. Thermal parameters fixed at $0.5\times10^4$ $pm^2$. Relative occupancies of planes fixed to reflect $Ca_{25}Co_{22}O_{56}(OH)_{28}$ formula: $(0.573CaO)_2CoO_2$. OH is not included in the refinements.

[2]CaO plane height (Ca and O in the plane in a 1:1 ratio, at the same height) and O plane height (O above and below Co in the $CoO_2$ layer) both at z and −z in the cell, Co fixed at z=0: Two structural parameters refined.

**Figure Captions**

**Fig.1.** [001] electron diffraction pattern of $Ca_{25}Co_{22}O_{56}(OH)_{28}$ showing the hexagonal and square layer stacking and showing the misfit character of the structure. In the upper right corner, the reciprocal lattices corresponding to the hexagonal and rock salt like parts of the structure are indicated by dashed and full lines respectively. In addition to the diffraction spots due to these two types of layers, extra spots can be seen due to the "beating" of both sublattices.

**Fig. 2.** Model of the misfit character of $Ca_{25}Co_{22}O_{56}(OH)_{28}$. (a) The upper part of the figure shows a square CaO layer and the lower part of the figure shows the oxygens in the adjacent plane that are part of the neighboring hexagonal $CoO_2$ layer. (b) Model for the structure of $Ca_{25}Co_{22}O_{56}(OH)_{28}$ showing two rocksalt layers and the $CoO_2$ plane – represented as $CoO_6$ octahedra sharing edges.

**Fig. 3**. The weight loss of a single phase sample of $Ca_{25}Co_{22}O_{56}(OH)_{28}$ on heating in 5% $H_2$ in Ar in a TGA at a rate of 0.25 degrees C per minute to 800C. The products of the reduction are CaO and Co metal.

**Fig.4.** The powder X-ray diffraction pattern (Log scale, Cu Kα radiation) of the sample with composition $Ca_{25}Co_{22}O_{56}(OH)_{28}$.

**Fig.5.** Inverse of magnetic susceptibility $(1/(\chi-\chi_0))$ vs. temperature of $Ca_{25}Co_{22}O_{56}(OH)_{28}$. The line represents the Curie-Weiss fit. Inset: temperature variation of the magnetic susceptibility.

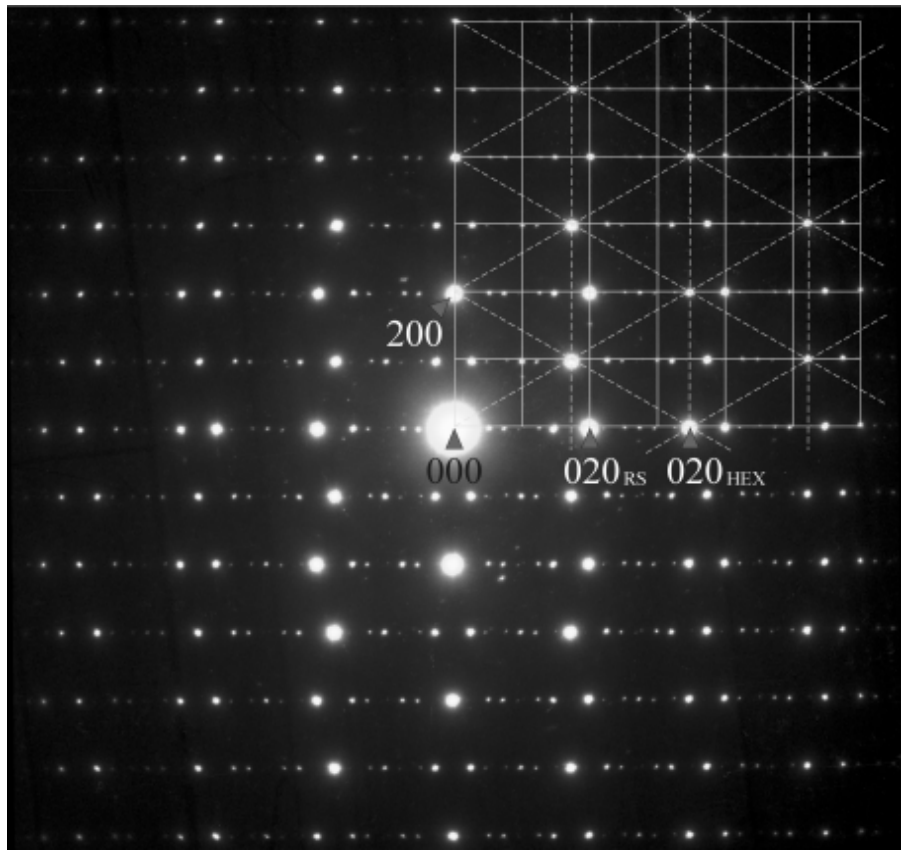

Figure 1

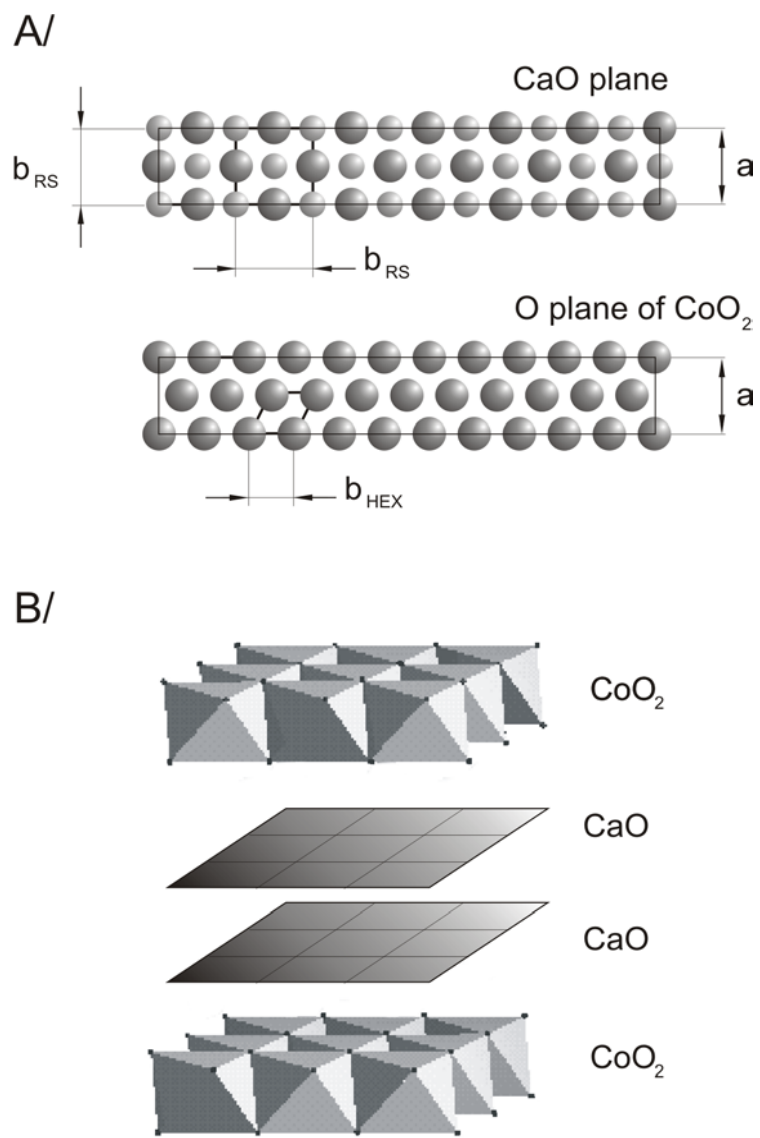

Figure 2

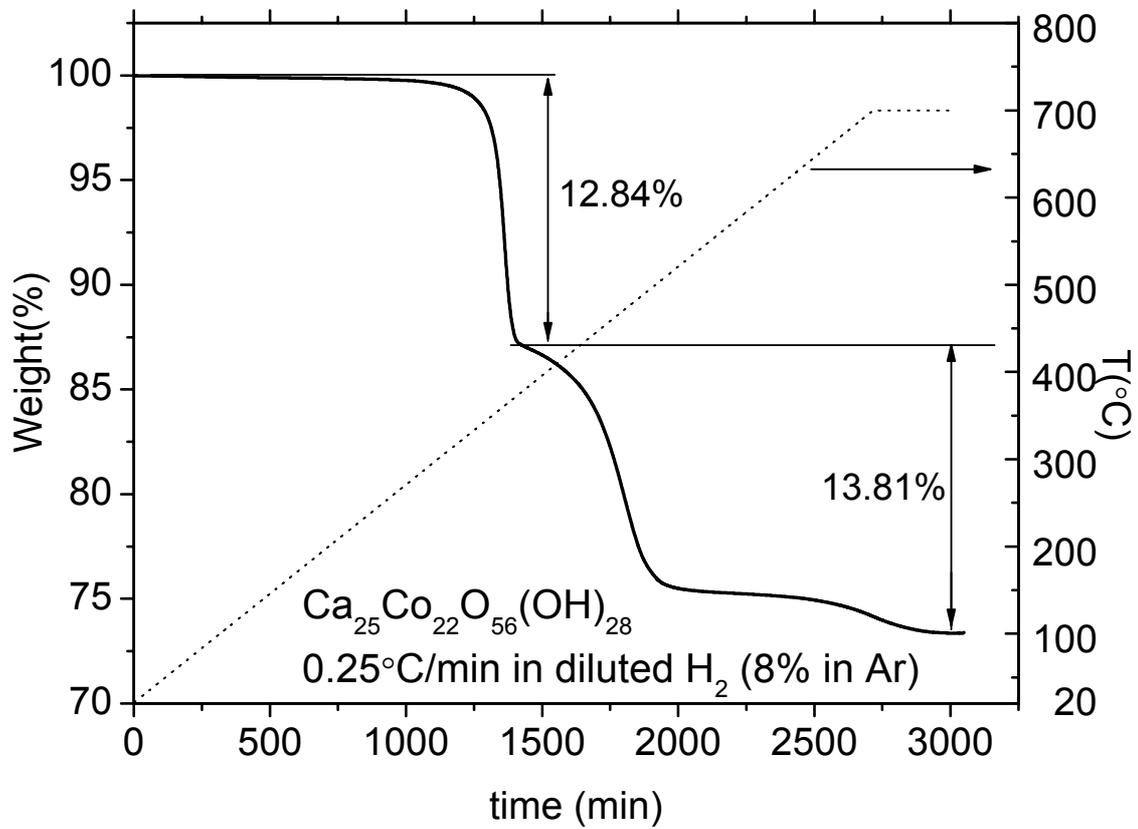

Figure 3

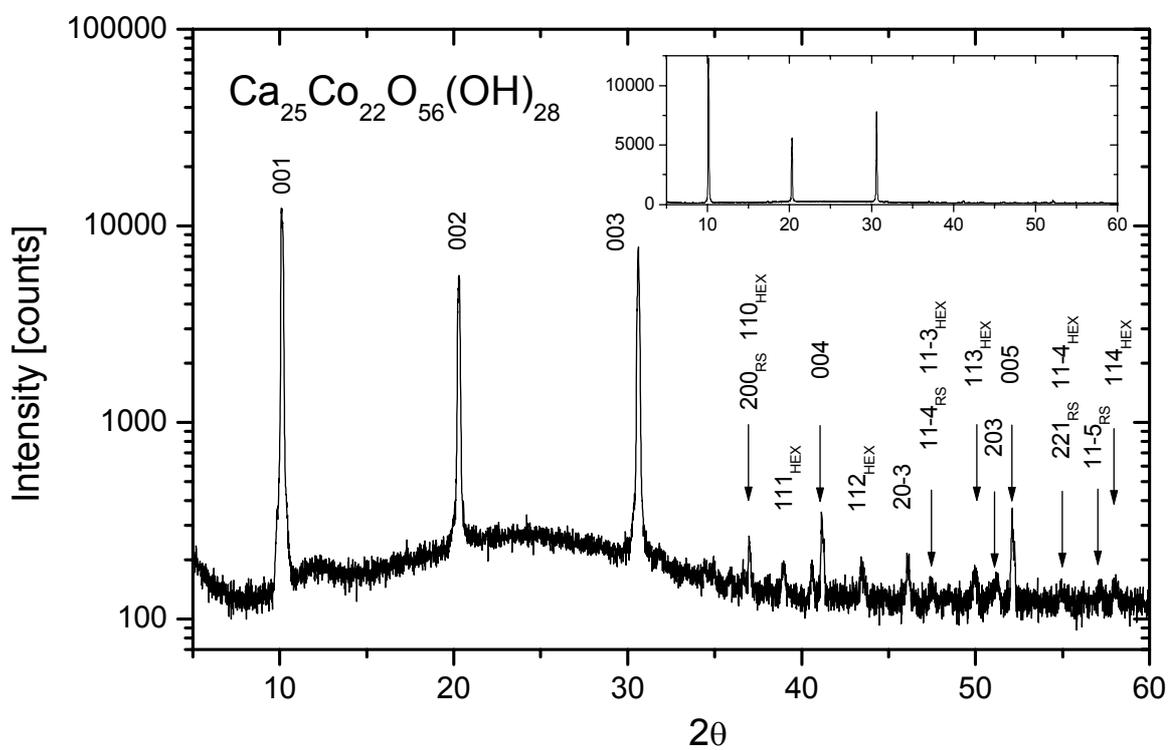

Figure 4

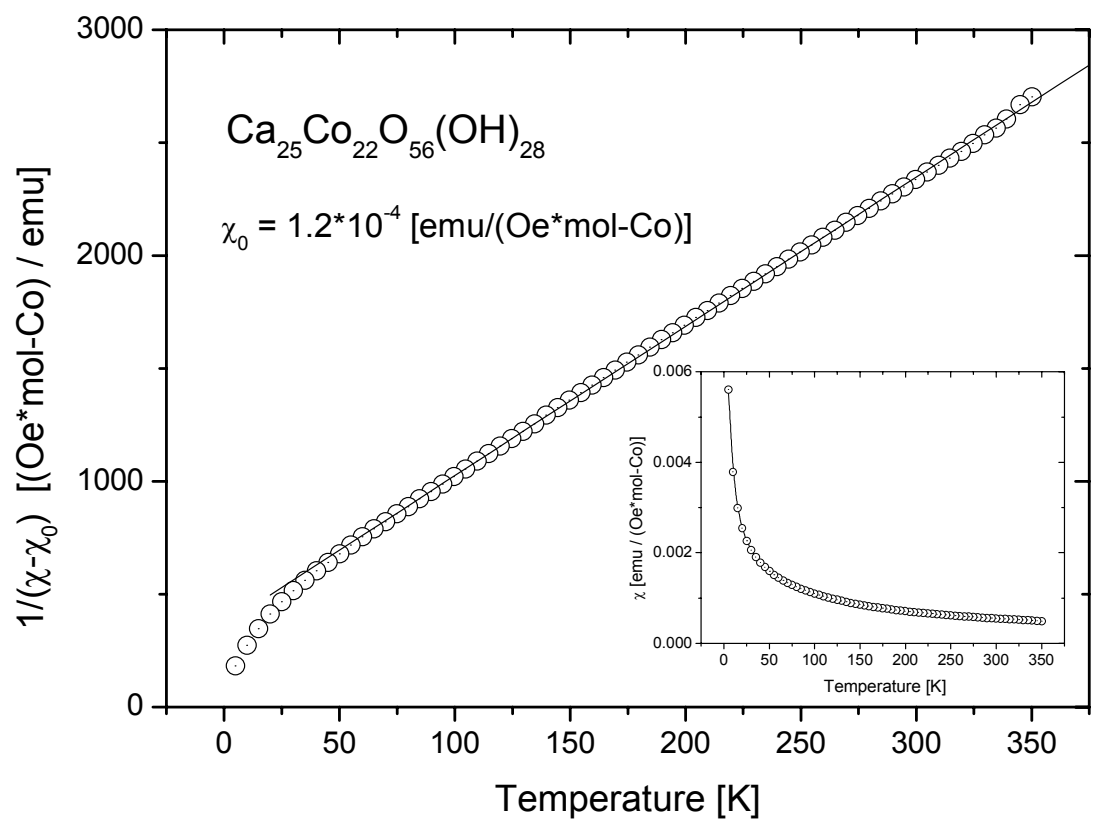

Figure 5